\documentstyle[12pt]{article}
\begin{document}
\begin{center}
UNIFIED STUDY OF PLANAR FIELD THEORIES\\
\vskip 5cm
Subir Ghosh\footnote{email:
  sghosh@isical.ac.in ; subir@boson.bose.res.in}\\
\vskip 2cm
P. A. M. U., Indian Statistical Institute,\\
203 B. T. Road, Calcutta 700035, India.
\end{center}
\vskip 4cm
Abstract:\\
\vskip .5cm
A "Master" gauge theory is constructed in 2+1-dimensions through which
various gauge invariant and gauge non-invariant theories can be studied.
In particular, Maxwell-Chern-Simons, Maxwell-Proca and Maxwell-Chern-Simons
-Proca models are considered here. The Master theory in an enlarged phase
space is constructed both in Lagrangian (Stuckelberg) and 
 Hamiltonian (Batalin-Tyutin) frameworks,
 the latter being the more general one, which includes the former as a
 special case. 
 Subsequently, BRST quantization of the latter is performed. Lastly, the
master Lagrangian, constructed by Deser and Jackiw 
(Phys. Lett. B139, (1984)  371), to show the equivalence between the 
Maxwell-Chern-Simons and the 
self-dual model, is also reproduced from our Batalin-Tyutin extended model.   
 Symplectic quantization procedure for constraint systems is adopted 
in the last demonstration.

\newpage
\begin{center}
{\bf I: Introduction}
\end{center}
\vskip 1cm
The Maxwell-Chern-Simons (MCS) theory allows the presence of massive modes
in the gauge field sector without breaking gauge invariance. This is 
possible because of the existance of the topological Chern-Simons (CS)
 term in
2+1-dimensions. This has been demonstrated long ago by Schonfeld \cite{sc}
 and by Deser,
Jackiw and Templeton \cite{djt}. The two First Class Constraints (FCC)
(according to the classification scheme of Dirac \cite{di}) in the theory
removes two degrees of freedom. The violation of parity, induced by the
CS term, is manifested in the appearance of a single helicity
component of the remaining excitation, the sign of which is correlated
with that of the coefficient
of the CS term.

However, in the MCS model, one can include a conventional vector field mass
term (or Proca term)
 which explicitly breaks gauge invariance. The resulting model, known as
 the MCS-Proca (MCSP) model, was analysed by 
Pisarski and Rao at the perturbative one loop level and by
Paul and Khare \cite{pk} in
 the Lagrangian framework. Later it was shown \cite{sg1} that this model
 appears naturally, under certain approximations, in the recently studied
 higher dimensional bosonization \cite{fs}.

  The MCSP model was considered
 briefly in \cite{des} in the Lagrangian framework. The detailed constraint
 analysis in the Hamiltonian framework was carried out in \cite{sg2}.

 The effect of the Proca term is quite interesting. It breaks the gauge
 invariance resulting in a theory with two Second Class Constraints (SCC)
 \cite{di},
 instead of the two FCCs present before. This change brings to life a second
 massive mode. The parity violation is seen in the fact that the above
 mentioned two modes of {\it distinct} masses carry spins $\pm 1$. This is
 similar to the Maxwell-Proca (MP) model, the difference being that here parity
 is not broken, which is reflected in the {\it same} masses of the opposite
 spins. Recently it was directly established
 \cite {bkm} that MCSP theory is a free one,
 comprising of the modes discussed above, via a complicated set of
 canonical transformations.

Let us now put the present work in its proper perspective.
From the above introduction, it is evident that both the gauge invariant
and gauge non-invariant models in 2+1 dimensions have been studied mainly
in the Lagrangian formalism,
 (except \cite{bkm} 
where canonical transformations in a classical
Hamiltonian framework has been used). However, {\it all}
 the above models can be
conveniently discussed in a unified way,
in a Hamiltonian formulation, where different theories, such as
MP, MCS and MCSP, appear as special cases of an underlying "master" theory.
 This constitutes the main body of our work. 
The master theory constructed here is a gauge theory, which for different
choices of parameters and gauge fixing conditions, leads to the different
theories. Interestingly enough, we are also able to rederive the "master"
 Lagrangian proposed by Deser and Jackiw \cite{dj}, which demonstrated the 
equivalence between the MCS and self-dual model, as a special case.

This underlying theory is obtained from two different 
viewpoints: (i) the
Stuckelberg (Lagrangian) extension and (ii) BRST-BFV \cite{brs, bfv}
Hamiltonian scheme. In particular we follow a specific formulation of
the BFV scheme, known as the Batalin- Tyutin (BT)
\cite{bt} extension. Although in
spirit, both the formalisms, (i.e. Lagrangian and Hamiltonian),
 introduce extra dynamical degrees of freedom in
order to elevate a gauge non-invariant theory to a gauge invariant one, (in
the enlarged phase space), the BFV and BT \cite{bfv,bt} schemes are much more
general and are applicable for any kind of gauge breaking term. The
usefulness of this prescription, specially in non-linear theories has been
demonstrated in \cite{sg3}. Indeed, the Stuckelberg result follows as a
special case of the BT construction, as we will discuss below. Also,
being a gauge theory, these extended gauge models enjoy a wide range of
freedom, in the form of the choice of gauge fixing fermion, in 
BRST quantization.

    Recent applications of these
field theoretic models in condensed matter systems, where the dynamics
 normal to a plane is severely restricted, have proved very fruitful
 \cite{frd}.

The plan of the paper is the following: In section II the Stuckelberg
construction is discussed. The existing results corresponding to the
different models are rederived in a unifed maner. The BT extension
is developed in section III, where the set of FCCs, the First Class (FC)
variables as well as the FC Hamiltonian in involution are constructed.
Section IV deals with the path integral quantization of the BT extended
model, in the BRST framework. The connection with the Stuckelberg
Lagrangian is also shown here. The emergence of the master lagrangian in
\cite{dj} from our BT extension is demonstrated 
in section V. The Deser-Jackiw model is analysed 
in the Faddeev-Jackiw formalism 
\cite{fj}, which is more convenient in the present case.
 The paper ends with a conclusion in
section VI.

\vskip 1cm
\begin{center}
{\bf II: Stuckelberg extension}
\end{center}

The MCSP model, with the metric being $g_{\mu\nu }=diag(+--),
~\epsilon _{12}=1$, is
\begin{equation}
{\cal L}_{MCSP}=-{1\over 4}A_{\mu\nu }A^{\mu\nu }+{{\mu }\over 4}
\epsilon_{\mu\nu\lambda }A^{\mu\nu }A^\lambda +{{m^2}\over 2}A_\mu A^\mu ,
\label{j}
\end{equation}
where $A_{\mu \nu }=\partial _\mu A_\nu-\partial _\nu A_\mu $.
Taking $m^2=0$ reproduces the MCS theory, which being a gauge theory is
 amenable to gauge fixing conditions. This simplifies the model considerably
  and makes the field content transparant. In order to discuss the MCS
theory as well, let us now construct the 
 master Lagrangian by converting the above gauge
    non-invariant theory to a gauge invariant one by the Stuckelberg
     prescription, For this purpose, we introduce the Stuckelberg field
     $\theta $ and define the extended Lagrangian as,

\begin{equation}
{\cal L}_{St}=-{1\over 4}A_{\mu\nu }A^{\mu\nu }+{{\mu }\over 4}
\epsilon_{\mu\nu\lambda }A^{\mu\nu }A^\lambda
+{{m^2}\over 2}(A_\mu -\partial _\mu \theta )(A^\mu -\partial ^\mu \theta ).
\label{k}
\end{equation}
In the ensuing gauge theory, we define the conjugate
 momenta \cite{djt} and the Poisson bracket algebra as,
$$
{{\partial {\cal L}_{St}}\over {\partial \dot A^i}}\equiv \Pi ^i=-\dot
 A_i+\partial _iA_0-{\mu \over 2}\epsilon_{ij}A_j;
~~{{\partial {\cal L}_{St}}\over {\partial \dot A^0}}\equiv \Pi ^0=m^2
\theta ;~~{{\partial {\cal L}_{St}}\over {\partial \dot \theta }}\equiv
 \Pi _\theta =m^2\dot \theta ,$$
\begin{equation}
\{A_\mu (x),\Pi _\nu (y)\}=-g _{\mu \nu }\delta (x-y),~ \{\theta (x),
\Pi _\theta (y)\}=\delta (x-y) .
\label{l}
\end{equation}
The Hamiltonian is
$$
{\cal H}_{St}=\Pi ^\mu \dot A^\mu + \Pi _\theta \dot \theta -{\cal L}_{St} $$
$$
={1\over 2}\Pi _i^2+{1\over 4}A_{ij}A_{ij}+({{m^2}\over 2}
+{{\mu ^2}\over 8})A_i A_i
-{\mu \over 2}\epsilon_{ij}\Pi _i A_j $$
\begin{equation}
+{1\over {2m^2}}\Pi _\theta ^2
 +{{m^2}\over 2}\partial _i\theta \partial _i\theta +m^2(\partial _iA_i
 )\theta
-A_0(\partial _i\Pi _i+{\mu \over 2}\epsilon _{ij}\partial _iA_j+{{m^2}
\over 2}A_0),
\label{mm}
\end{equation}
where a total derivative term has been dropped. The two 
FCCs in involution are,
\begin{equation}
\chi _1\equiv \Pi _0-m^2\theta,~~\chi _2\equiv \partial _i\Pi _i+{\mu \over 2}\epsilon _{ij}\partial _iA_j+m^2 A_0 +\Pi _\theta .
\label{n}
\end{equation}
 The unitary gauge,
$\eta ^1\equiv \Pi _\theta;~~\eta ^2\equiv \theta $,
establishes gauge equivalence between the embedded model and the original
 MCSP model. This ensures that in the gauge invariant sector, results
  obtained in any convenient gauge will be true for the MCSP theory.
We invoke the rotationally symmetric Coulomb gauge \cite{djt}
\begin{equation}
\eta ^1\equiv A_0;~~~\eta ^2\equiv \partial _i A_i .
\label{o}
\end{equation}
The $(\chi _i,~\eta ^j )$ system of four constraints are now second class,
 meaning that the constraint algebra 
matrix is invertible. The Dirac brackets, defined in the conventional way,
 are given below:
$$
\{A_i(x),\Pi _j(y)\}^*=(\delta _{ij}-{{\partial _i\partial _j}\over
 {\nabla ^2}})\delta(x-y);~~\{\Pi_i(x),\Pi_j(y)\}^*
=-{\mu \over 2}\epsilon _{ij}\delta (x-y) $$
\begin{equation}
\{\Pi _i(x),\theta (y)\}^*={{\partial _i}\over {\nabla ^2}}\delta(x-y);
~~\{\Pi _i(x),\Pi _0 (y)\}^*=-m^2{{\partial _i}\over {\nabla ^2}}\delta(x-y).
\label{p}
\end{equation}
The remaining brackets are same as the Poisson brackets. The reduced
 Hamiltonian in Coulomb gauge is
$$
{\cal H}_S={1\over 2}\Pi _i^2+{1\over 2}\partial _iA_j\partial _iA_j
+({{m^2}\over 2}+{{\mu ^2}\over 8})A_i A_i
-{\mu \over 2}\epsilon_{ij}\Pi _i A_j
$$
\begin{equation}
+
{1\over {2m^2}}\Pi _\theta ^2
 +{{m^2}\over 2}\partial _i\theta \partial _i\theta .
\label{q}
\end{equation}
Although somewhat tedious, it is straightforward to verify that the
 following combinations,
$\phi =((\epsilon _{ij}\partial _i A_j),(\epsilon _{ij}\partial _i\Pi _j)
,\Pi _\theta , \theta )$ obey the higher derivative equation
\begin{equation}
(\Box +M_1^2 )(\Box +M_2^2)\phi =0;~~M_1^2(M_2^2)={1\over 2}[2m^2+\mu ^2\pm
 \mu {\sqrt {\mu ^2+4m^2}}].
\label{u}
\end{equation}
The spectra agrees with \cite{pk,sg2}. Note that
for $\mu ^2 =0$, the roots collapse to $M_1^2=M_2^2=m^2$, which is just the
 Maxwell-Proca model, whereas for
$m^2=0 $, in MCS theory, 
 the roots are $M_1^2=\mu ^2,~M_2^2=0 $, indicating the presence of
 only the topologically massive mode, since the Stuckelberg field $\theta
  $ is no longer present.

Prior to fixing the $\eta _2 $ gauge, the gauge invariant sector is
 identified as,
\begin{equation}
E_i=-\Pi _i+{\mu \over 2}\epsilon _{ij}A_j;~~ B=-\epsilon _{ij}
\partial _iA_j;~~\Pi _\theta ;~~A_i+\partial _i\theta ,
\label{y}
\end{equation}
where $E_i$ and $B$ are the conventional electric and magnetic fields.
In the reduced space, the Hamiltonian and spatial translation generators
 are gauge invariant,
$${\cal H}_{St}={1\over 2}(E_i^2+B^2+{{\Pi _\theta ^2}\over {m^2}}+m^2(A_i+
\partial _i\theta )^2),$$
\begin{equation}
{\cal P}_{St}^i=-\epsilon _{ij}E_jB-\Pi _\theta (A_i+\partial _i\theta ).
\label{x}
\end{equation}
Defining the boost transformation as $M^{i0}=-t\int d^2x{\cal P}_{St}^i(x)+
\int d^2xx^i{\cal H}_{St}(x)$, the Dirac brackets with the gauge invariant
 variables are easily computed. They will contain non-canonical pieces in
  order to be consistent with the constraints. However, changing to a new
   set of variables by the following canonical transformations,
\begin{equation}
Q_1(Q_2)={1\over {{\sqrt {-2\nabla ^2}}}}[\epsilon _{ij}\partial _iA_
j\pm {1\over m}\Pi _\theta ];~~
P_1(P_2)=[{1\over {{\sqrt {-2\nabla ^2}}}}\epsilon _{ij}\partial _i\Pi _j
\mp
{m\over 2}{\sqrt {-2\nabla ^2}}\theta ],
\label{w}
\end{equation}
we can convert our system to a nearly decoupled one.
Passing on to the quantum theory, the redefined variables satisfy the
 canonical algebra,
\begin{equation}
i[P_i,Q_j]=\delta _{ij}\delta (x-y);~~[Q_i,Q_j]=[P_i,P_j]=0;
~~i,j=1,2~.
\label{xy}
\end{equation}
The electric and magnetic fields and the translation generators are
 rewritten as,
\begin{equation}
B=-{{\sqrt {-2\nabla ^2}}\over 2}(Q_1+Q_2);~~
E_i=-{1\over {\sqrt {-2\nabla ^2}}}[\epsilon _{ij}\partial _j(P_1+P_2)
+(\mu
 + m)\partial _iQ_1+(\mu -m)\partial _iQ_2],
\label{z}
\end{equation}
$$
H_{St}=\int d^2x[{1\over 2}(P_1^2+\partial _iQ_1\partial _iQ_1+M_1^2Q_1^2)
+{1\over 2}(P_2^2+\partial _iQ_2\partial _iQ_2+M_2^2Q_2^2)+{{\mu ^2}\over 2}Q_1Q_2]$$
\begin{equation}
P^i_{St}=\int d^2x[P_1\partial ^iQ_1+P_2\partial ^iQ_2]
\label{ff}
\end{equation}
It is worthwhile to 
 consider the special cases, $m^2=0$ or $\mu ^2=0$
 corresponding to MCS and MP models respectively. In the former choice,
   {\it i.e.}  in the MCS theory, as we noted before, $\theta $ field 
is absent, which makes the $(Q_1,~P_1)$ pair identical to the
 $(Q_2,~P_2)$ pair,
    leading to the following relations, with $i[p(x),q(y)]=\delta (x-y)$,
$$B={\sqrt{-\nabla ^2}}q,~~E_i={1\over {{\sqrt{-\nabla ^2}}}}(\epsilon_{ij}
\partial _jp+\mu\partial _iq),$$
\begin{equation}
H=\int d^2x {1\over 2}(p^2+\partial _iq\partial _iq+\mu ^2q^2),~~P^i=\int
 d^2x(p\partial ^iq).
\label{mcs}
\end{equation}
This set of relations is identical to those in \cite{djt} and hence their
 result, that the spin is $\pm \mu /\mid \mu \mid $, will follow in a
  straightforward fashion. Due to parity violation, depending on the sign
   of $\mu $, only one component of spin is present.

The latter case, $\mu ^2=0$, refers to the Proca model, where $M_1^2=M_2^2
=m^2$, and we get,
$$
B=-{{\sqrt {-2\nabla ^2}}\over 2}(Q_1+Q_2);~~
E_i=-{1\over {\sqrt {-2\nabla ^2}}}[\epsilon _{ij}\partial _j(P_1+P_2)
+m(\partial _iQ_1 -\partial _iQ_2)],
$$
$$
H=\int d^2x[{1\over 2}(P_1^2+\partial _iQ_1\partial _iQ_1+m_1^2Q_1^2)
+{1\over 2}(P_2^2+\partial _iQ_2\partial _iQ_2+m_2^2Q_2^2)],$$
\begin{equation}
P^i=\int d^2x[P_1\partial ^iQ_1+P_2\partial ^iQ_2].
\label{fg}
\end{equation}

Let us briefly outline the analysis of DJT \cite{djt} where the subtle
 interplay between Poincare invariance and an unambiguous determination
  of the spin of the excitations in a vector theory was revealed. It was
   shown that the correct space-time transformation of the gauge invariant
    observables, such as electric and magnetic fields, were induced by
     Poincare generators which obeyed an anomalous algebra among themselves.
      However, a phase redefinition of the creation and annihilation
       operators removed the commutator anomaly and yielded the spin
        contribution in a single stroke.

Following the prescription of DJT given in \cite{djt}, the boost generator
 $M^{i0}$ should be reinforced by the additional terms,
$$
m\epsilon_{ij}\int d^2x({{P_1\partial _jQ_1}\over{-\nabla ^2}}
-{{P_2\partial _jQ_2}\over{-\nabla ^2}}),$$
such that the electromagnetic fields transform correctly. This addition,
 however, generates a zero momentum anomaly in the boost algebra,
\begin{equation}
i[M^{i0},M^{j0}]=\epsilon^{ij}(M-\Delta),~~~\Delta={{m^3}\over{4\pi }}
\{(\int Q_1)^2-(\int Q_2)^2\}+{m\over{4\pi }}\{(\int P_1)^2-(\int P_2)^2\},
\label{pr}
\end{equation}
where $M$ is the rotation generator 
$$M=-\int d^2x(P_1\epsilon^{ij}x^i\partial _jQ_1+P_2\epsilon^{ij}x^i
\partial _jQ_2).$$
Making the mode expansions,
\begin{equation}
Q_1(x)(Q_2(x))=\int {{d^2k}\over{2\pi {\sqrt{2\omega (k)}}}
}[e^{-ikx}a(k)(b(k))+e^{ikx}a^+(k)(b^+(k))],
\label{mo}
\end{equation}
and effecting the phase redefinitions,
\begin{equation}
a\rightarrow e^{i{m\over{\mid m\mid }}\theta }a,~~~b\rightarrow
 e^{-i{m\over{\mid m\mid }}\theta }b,
\label{ph}
\end{equation}
where $\theta =tan^{-1}k_2/k_1 $, one recovers the full angular momentum
 as
\begin{equation}
M=\int d^2k(a^+(k){1\over i}{\partial \over {\partial \theta }}a(k)+
b^+(k){1\over i}{\partial \over {\partial \theta }}b(k))+{m\over {\mid m\mid }}\int d^2k(a^+(k)a(k)-b^+(k)b(k)),
\label{sp}
\end{equation}
where the second term is the spin. This indicates parity non-violation as
 opposite spins are contributed by modes having the same mass.

Let us now come to MCSP model. In this case we choose a gauge of the
form
\begin{equation}
\eta ^1\equiv A_0~;~~\eta ^2\equiv \theta .
\label{gauge}
\end{equation}
We can now implement the four SCCs strongly in the Hamiltonian by using
$$A_0=\theta =0~~;~~ \Pi_\theta =-(\partial_i\Pi_i+
{\mu \over 2}\epsilon_{ij}\partial_iA_j).
$$
The Hamiltonian simplifies to,
$$
{\cal H}
={1\over 2}\Pi _i^2+{1\over 4}A_{ij}A_{ij}+({{m^2}\over 2}
+{{\mu ^2}\over 8})A_i A_i 
$$
\begin{equation}
-{\mu \over 2}\epsilon_{ij}\Pi _i A_j 
+{1\over {2m^2}}(\partial_i\Pi_i+{\mu \over 2}\epsilon_{ij}\partial_iA_j)^2.
\label{nn}
\end{equation}
Notice that this choice does not affect the symplectic structure of the
remaining variables. 
{\footnote {Effectively this gauge is same as the conventional
unitary gauge, (where one chooses BT variables to be zero as the gauge
choice), as far as the present analysis is concerned.}}
 The following
canonical transformations discussed
in \cite{bkm} are applicable here,
$$A_i={{2m}\over{{\sqrt {4m^2+\theta ^2}}}}\epsilon_{ij}{{\partial_j(Q_1+Q_2)}
\over{{\sqrt {-\nabla ^2}}}}+{1\over{2m}}{{\partial_i(P_1-P_2)}
\over{{\sqrt {-\nabla ^2}}}},$$
\begin{equation}
\Pi_i={{{\sqrt {4m^2+\theta ^2}}}\over{4m}}\epsilon_{ij}
{{\partial_j(P_1+P_2)}
\over{{\sqrt {-\nabla ^2}}}}-m{{\partial_i(Q_1-Q_2)}
\over{{\sqrt {-\nabla ^2}}}},
\label{ctr}
\end{equation}
and the decoupled Hamiltonian is
\begin{equation}
H=\int d^2x[{1\over 2}(P_1^2+\partial _iQ_1\partial _iQ_1+M_1^2Q_1^2)
+{1\over 2}(P_2^2+\partial _iQ_2\partial _iQ_2+M_2^2Q_2^2)],
\label{hh}
\end{equation}
(It should be noted that we have checked quite exhaustively to conclude
 that
any Coulomb like gauge is {\it not} able to lead to this decoupling.) The
resulting mass spectra and spin are discussed thoroughly in \cite{bkm}
and is not repeated here. 
 Substituting the transformations (\ref{ctr}) in the boost generator,
one can see \cite{bkm} that the anomalous term is a decoupled sum of
$'1'$ and $'2'$ variables and 
 incorporating phase redefinition as in (\ref{ph}) with
$m$ replaced by $M_1$ and $M_2$, one can recover the analogue of (\ref{sp})
with $M_1$ and $M_2$ for $a$ and $b$ operators respectively.
  Now
parity violation is manifest as the opposite helicities carry {\it different}
masses.
\vskip 1cm
\begin{center}
{\bf III:  Batalin-Tyutin extention}
\end{center}
In this section we will discuss the
 other (Hamiltonian) alternative, i.e. the BT extension, which is required
to convert the SCCs to FCCs since quantization of a gauge theory is
more familiar to us. (Also, in general, presence of SCCs can complicate
the symplectic structure and the path integral measure.) The idea of
enlarging the phase space in Hamiltonian framework, to be considered here,
is similar in spirit to the Stuckelberg extension, which is in Lagrangian
framework. However, the advantage of the former is that it can be
 applied \cite{sg3} in complicated
non-linear SC systems as well. In fact, eventually we will show 
that, for
simple systems, the latter appears as a special case of the former.

The explicit expressions regarding the additional terms, depending on
the BT degrees of freedom, required by the SCCs for conversion to FCCs
are provided in \cite{bt}.
 Hence
we simply show the results. From the original SCCs we derive the
commuting FCCs as
\begin{equation}
\tilde\chi_1\equiv \chi_1 -m^2\psi~;~~\tilde \chi_2\equiv \chi_2+\Pi_\psi,
\label{fcc}
\end{equation}
where the conjugate pair, 
$\{\psi (x), \Pi_\psi (y) \}=\delta(x-y)$, are the BT fields. 
 One can also introduce
a useful set of FC variables \cite{bt} that commute with $\tilde\chi_i$ by
construction. Once again exploiting the formulas given in \cite{bt},
 these are computed as,
$$\tilde {A}_0=A_0+{1\over{m^2}}\Pi_\psi~;~\tilde {A}_i=A_i+\partial_i\psi~;$$
\begin{equation}
\tilde\Pi_0=\Pi_0-m^2\psi ~;~\tilde\Pi_i=\Pi_i+{{\mu }\over 2}\epsilon_{ij}
\partial_j\psi .
\label{fcv}
\end{equation}
Next the Hamiltonian is rewritten in terms of the FC fields as,
$$
\tilde{\cal H}= 
{1\over 2}(\Pi _i+{\mu \over 2}\epsilon_{ij}\partial_j\psi )^2
+{1\over 4}{\tilde A}_{ij}{\tilde A}_{ij}$$
$$
+({{m^2}\over 2}
+{{\mu ^2}\over 8})(A_i +\partial_i\psi )^2
-{\mu \over 2}\epsilon_{ij}(\Pi _i+{\mu \over 2}\epsilon_{ik}\partial_k\psi )
(A_j+\partial_j\psi ) 
+{{m^2}\over 2}(A_0+{1\over {m^2}}\Pi_\psi ) ^2 $$
$$=
{1\over 2}(\Pi _i )^2
+{1\over 4} A_{ij} A_{ij}+({{m^2}\over 2}
+{{\mu ^2}\over 8})(A_i )^2
-{\mu \over 2}\epsilon_{ij}\Pi _iA_j +{{m^2}\over 2}A_0^2$$
\begin{equation}
+{1\over{2m^2}}\Pi _\psi ^2+A_0\Pi _\psi +{{m^2}\over 2}((\partial _i\psi ) ^2
+2A_i\partial_i\psi ).
\label{fch}
\end{equation}
It can be proved \cite{bt} that $\tilde {\cal {H}}$ commutes with 
$\tilde {\chi } _i$, by construction.
Notice tht we have dropped the term proportional to $\tilde\chi_2$ from
the Hamiltonian since the "constraint terms" coupled to arbitrary
multiplier fields will eventually appear in the action for BRST
quantization. The original SCCs, written in terms of the FC variables are
identical to the modified FCCs. This completes the BT extension.
\vskip 1cm
\begin{center}
{\bf IV: BRST quantization}
\end{center}
\vskip .5cm
The BRST quantization 
\cite{brs, bt, sg3}, in the enlarged phase space,  
 proceeds in the conventional way since only FCCs are
present. The phase space is further extended by introducing ghost,
 anti-ghost and multiplier fields. These are,
 the canonically conjugate fermionic ghost and anti-ghost pairs  
$(C^i(x),\bar P_i(x))$ and  $(P^i(x),\bar C_i(x))$ respectively
and the bosonic
multipliers and their  momenta   $(q^i(x),p_i(x))$.
The BRST charge $Q_{BRST}$ is defined as
$$ Q_{BRST}=\int d^2x (C^i\tilde\chi _i+P^ip_i) 
~;~  \{Q,Q\}=0~;~\{H_{BRST},Q_{BRST}\}=0.$$
In the present 
instance $\tilde{\cal H}={\cal H}_{BRST}$ since the system is
completely abelian with the FCCs and $\tilde {\cal H}$ being strictly in
involution. The unitary Hamiltonian,
$$H_U\equiv H_{BRST}-\{\Psi,Q_{BRST}\}=\tilde H-\{\Psi,Q_{BRST}\}$$
is obtained as,
\begin{equation}
H_U=\tilde H-[\bar P_iP^i+\bar C_i\{\eta ^i,\tilde \chi _j\}C^j+p_i\eta ^i
+q^i\tilde\chi _i].
 \label{hu}
\end{equation}
 The gauge fixing
fermion operator $\Psi $ is defined as,
$$\Psi =\int d^2x(\bar C_i\eta ^i+\bar P_iq^i),$$
where $\eta ^i$ are arbitrary functions of the fields, which can be 
considered as conventional gauge fixing conditions. The Hamiltonian
path integral is,
$$
Z=\int {\cal D}[\alpha]exp^{iS_{BRST}}~;~{\cal D}[\alpha]={\cal D}[A_\mu ,
\Pi _\mu, \psi ,\Pi _\psi ,C^i,\bar P_i,P^i,\bar C_i, q^i,p_i]~;$$
\begin{equation}
{\cal L}_{BRST}=\dot A_\mu 
\Pi ^\mu + \dot\psi \Pi _\psi +\bar P_i\dot C^i+\bar C_i\dot P^i
+\dot q^ip_i -{\cal H}_U .
\label{lbr}
\end{equation}
The sector of physical states is defined by 
$$Q_{BRST}\mid Ph>=0.$$
With the Poisson brackets replaced by commutators or anti-commutators, 
this describes the BRST quantization. Now, according to our motivation,
this path integral can be simplified further by 
 choosing specific forms of the arbitrary
functions $\eta ^i$, which actually amounts to a gauge fixing in the 
space of physical degrees of freedom. The only restriction 
 on this choice is that the
total system of constraints $(\tilde \chi _i, \eta ^j)$ must have a
non-vanishing Poisson bracket.

In order to forge a connection with the Stuckelberg construction, (as stated
before), let us consider a Coulomb like gauge,
\begin{equation}
\eta ^1\equiv A_0~;~\eta ^2\equiv -\partial _iA_i+\sigma (x),
\label{gg}
\end{equation}
where $\sigma $ is an undetermined, scalar field function. Since
$$\int d^2x(p_1\dot q^1+\bar C_1\dot P^1)=\{Q_{BRST},\int d^2x~\bar C_1
\dot q^1\}$$
the terms on the left hand side are removed from ${\cal L}_{BRST}$
 because $\bar C_1
\dot q^1 $ in the right hand side can be absorbed in the arbitrary
 $\eta ^1$ term. 
Integrating out $P^1$, $\bar P_1$ and $p_1$ results in $\delta (A_0)$ in
the measure, which removes the $A_0$ integral. Similarly, $q^1$ and
$\Pi_0 $ are trivially integrated out. Next, $\bar P_2$ and $P^2$ are
integrated leading to the relation $P^2=\dot C^2$. Hence we are left with
$$Z=\int {\cal D}[\alpha]exp^{iS_{BRST}}~;~ 
{\cal D}[\alpha]={\cal D}[A_\mu ,
\Pi _\mu, \psi ,\Pi _\psi ,C^i,\bar C_i, q^2,p_2],$$
$${\cal L}_{BRST}=\dot A^i 
\Pi ^i + \dot\psi \Pi _\psi 
+\dot q^2p_2 -p_2\eta ^2-q^2(\partial _i\Pi _i+{\mu\over 2}\epsilon_{ij}
\partial _iA_j +\Pi _\psi )
$$
$$ +\bar C_2\ddot C^2-\bar C_1
\{\eta ^1,\bar\chi _1\}C^1 -\bar C_2
\{\eta ^2,\bar\chi _2\}C^2
$$
\begin{equation}
-[{1\over 2}\Pi _i^2+{1\over 4}A_{ij}A_{ij}+({{m^2}\over 2}
+{{\mu ^2}\over 8})A_i A_i -{\mu \over 2}\epsilon_{ij}\Pi _i A_j 
+{1\over {2m^2}}\Pi _\psi ^2+{{m^2}\over 2}(\partial _i\psi \partial _i\psi
+2A_i\partial _i\psi)].
\label{01}
\end{equation}
For the gauge conditions that we have chosen, we get,
$$\{\eta ^1(x),\bar\chi _1(y)\}=-\delta(x-y)~;~
\{\eta ^2(x),\bar\chi _2(y)\}=\nabla ^2\delta (x-y)
+\{\sigma(x),\bar\chi _2(y)\}.$$
Hence the ghost part of the Lagrangian becomes,
$$\bar C_1C^1-\partial _\mu\bar C_2\partial ^\mu C^2 
-\bar C_2
\{\sigma ,\bar\chi _2\}C^2.$$
The gaussian integrals corresponding to $\Pi _i$ and $\Pi _\psi $
contribute respectively,
$${1\over 2}(\dot A_i+\partial _iq^2+{\mu \over 2}\epsilon_{ij}A_j)^2;~~~
{{m^2}\over 2}(\dot\psi -q^2)^2,$$
in the action.
Finally, identifying $q^2=-A_0$, (since $q^i$ were arbitrary anyway), and
combining all the above terms we obtain,
$$
Z=\int {\cal D}[\alpha]exp^{iS}~;~ 
{\cal D}[\alpha]={\cal D}[A_\mu ,
\psi ,\Pi _\psi ,\bar C, C]\delta(\partial _\mu A^\mu +\sigma ),$$
$$
{\cal L}=-{1\over 4}A_{\mu\nu }A^{\mu\nu }+{{\mu }\over 4}
\epsilon_{\mu\nu\lambda }A^{\mu\nu }A^\lambda +{{m^2}\over 2}A_\mu A^\mu $$
\begin{equation}
+{{m^2}\over 2}\partial _\mu \psi \partial ^\mu \psi +m^2A^\mu 
\partial _\mu \psi -\partial _\mu \bar C\partial ^\mu C
-\bar C\{\sigma ,\tilde \chi _2 \}C.
\label{lst}
\end{equation}
For $\sigma =0$, this is nothing but the Stuckelberg extension of the 
MCSP model, in the Lorentz gauge. This completes the identification of
the Stuckelberg extension with the Batalin-Tyutin extension for a
specific choice of the gauge fixing fermion in the latter scheme. In the
 next section, we provide a different gauge fixing fermion which leads
 to another interesting model.

\vskip 1cm
\begin{center}
{\bf V: Recovering the Deser-Jackiw "master" Lagrangian}
\end{center}
\vskip .5cm

The idea of constructing a "master" Lagrangian to show explicitly the
equivalence between various theories was pioneered by Deser and Jackiw
 \cite{dj}. The "master" Lagrangian is an interacting model from which
selective integration of some fields leads to the desired models and the
 latter models are termed as equivalent. It should be pointed out that
in general, integration of a dynamical field requires that the
 quantum effects be
considered. However, if the fields that are to be removed occur linearly
or quadratically, classical equations of motion suffice. In this sense,
 the Deser-Jackiw model \cite{dj} demonstrated the equivalence between
the abelian MCS theory and abelian self-dual theory. The "master"
 Lagrangian posited in \cite{dj} is,
$$
{\cal L}_{DJ}={1\over 2}f^\mu f_\mu-f_\mu \epsilon ^{\mu\alpha\beta}\partial
_\alpha A_\beta +{{\mu }\over 2}A_\mu \epsilon ^{\mu\alpha\beta}\partial
_\alpha A_\beta $$
\begin{equation}
={1\over 2}f^\mu f_\mu -\epsilon ^{ij}(f^0 \partial ^iA^j
-f^i \partial ^0A^j +f^i \partial ^jA^0) +{{\mu }\over 2}\epsilon ^{ij}
(2A^0 \partial ^iA^j-A^i \partial ^0A^j ).
\label{33}
\end{equation}
Since the Lagrangian is first order (in time derivatives), all the 
canonical momenta lead to primary constraints. They are,
\begin{equation}
P^\mu \equiv {{\partial {\cal L}_{DJ}}\over {\partial \dot f^\mu }}
\approx 0~~,
~~\Pi^0 \equiv {{\partial {\cal L}_{DJ}}\over {\partial \dot A^0 }}
\approx 0~~,
\Pi^j \equiv {{\partial {\cal L}_{DJ}}\over {\partial \dot A^j }}=
\epsilon^{ij}(f^i -{{\mu }\over 2}A^i).
\label{34}
\end{equation}
Clearly the proliferation of SCCs and the first order nature of the 
Lagrangian suggests that the Faddeev-Jackiw symplective quantization
scheme \cite{fj} would be more appropriate in the present case. In
 the Faddeev-Jackiw \cite{fj} formalism, one is allowed to bypass the
 derivation and classification of all the constraints, isolation of
 the FCCs and invoking the SCCs via introduction of Dirac brackets. 
 Instead, here a generic Lagrangian is expressed in the first order form as,
\begin{equation}
Ldt=a_id\rho ^i-V(\rho )dt,
\label{99}
\end{equation}
in which the symplectic structure is provided by,
\begin{equation}
\omega _{ij}={{\partial a_j}\over
{\partial \rho ^i}}-{{\partial a_i}\over
{\partial \rho ^j}}~~,~~\{\rho ^i(x),\rho ^j(y)\}=\omega _{ij}^{-1}(x,y),
\label{98}
\end{equation}
provided the matrix $\omega _{ij}$ is invertible. The aim is to express
 the Lagrangian in the following form,
\begin{equation}
Ldt=a_kd\bar\rho ^k-V(\bar\rho )dt -\lambda _l\Phi ^l(\bar\rho ),
\label{97}
\end{equation}
where $\bar\rho ^k$ are less in number than $\rho ^i$. Some of the latter
 variables appear as multiplier fields $\lambda _l$. $\Phi ^l$s are the
constraints of the theory.
The
Lagrangian in (\ref{33}) is rewritten as,
$${\cal L}_{DJ}=\epsilon ^{ij}({{\mu }\over 2}A^j-f^j)\dot A^i
+{1\over 2}f^\mu f_\mu-\epsilon ^{ij}[f^0 \partial ^iA^j-A^0
(\mu \partial ^iA^j+\partial ^jf^i)]$$
\begin{equation}
=\epsilon ^{ij}({{\mu }\over 2}A^j-f^j)\dot A^i
-{1\over 2}f^i f^i-{1\over 2}(\epsilon ^{ij}\partial ^iA^j)^2 +A^0
(\mu \epsilon ^{ij}\partial ^iA^j -\epsilon ^{ij}\partial ^if^j ),
\label{35}
\end{equation}
where the equation of motion for $f^0$, i.e. $f^0-\epsilon ^{ij}\partial ^iA^j=0$, has been used to
 eliminate $f^0$. $A^0 (\equiv \lambda )$ is simply a multiplier
field attached to the single FCC,
\begin{equation}
 \Phi \equiv \mu \epsilon ^{ij}\partial ^iA^j 
-\epsilon ^{ij}\partial ^if^j \approx 0.
\label{36}
\end{equation}
In order to obtain the symplectic structure of the remaining fields, we
rewrite the kinetic part of (\ref{35}) as
\begin{equation}
{\cal L}_{SD}^{symp}=\epsilon ^{ij}({{\mu }\over 2}A^j-f^j)\dot A^i
\equiv a_i\dot {\bar\rho }^i=a_3\dot {\bar\rho}^3+a_4\dot {\bar\rho}^4,
\label{37}
\end{equation}
where the following identifications are made,
$$\bar\rho ^1\equiv f^1~~,~~\bar\rho ^2\equiv f^2~~,~~\bar\rho ^3\equiv A^1~~,~~
\bar\rho ^4\equiv A^2~~,$$
$$a_1=a_2=0~~,~~
a_3=-\bar\rho ^2+{{\mu }\over 2}\bar\rho ^4~~,
~~a_4=-\bar\rho ^1+{{\mu }\over 2}\bar\rho ^3~~.$$
The symplectic two form matrix $\omega _{ij}$ is computed as,
$$
\displaylines{
\pmatrix{0&0&0&\delta (x-y)\cr
0&0&-\delta (x-y)&0\cr
0&\delta (x-y)&0&-\mu \delta (x-y)\cr
-\delta (x-y)&0&-\mu \delta (x-y)&0\cr}\cr}
$$
As the above matrix is invertible, one can directly read of the 
non-canonical symplectic structure from the definition,
\begin{equation}
\{\bar\rho ^i(x),\bar\rho ^j(y)\}=\omega _{ij}^{-1}(x,y),
\label{39}
\end{equation}
where the inverse matrix $\omega_{ij}^{-1}$ is,
$$
\displaylines{
\pmatrix{
0&-\mu \delta (x-y)&0&\delta (x-y)\cr
\mu \delta (x-y)&0&\delta (x-y)&0\cr
0&-\delta (x-y)&0&0\cr
\delta (x-y)&0&0&0\cr}\cr} 
$$
Returning to the original definition of variables, the new symplectic
 structure is,
\begin{equation}
\{f ^i(x),f^j(y)\}=-\mu \epsilon^{ij} \delta (x-y)~~,
~~\{f ^i(x),A^j(y)\}=-\epsilon ^{ij} \delta (x-y)~~,~~
\{A ^i(x),A^j(y)\}=0.
\label{40}
\end{equation}
Let us now go back to the $m^2=0$ limit of our BT 
 extended model in (\ref{fch}),
$${\cal L}_{BRST}=\dot A^i 
\Pi ^i  
-[{1\over 2}\Pi _i^2+{1\over 4}A_{ij}A_{ij}+
+{{\mu ^2}\over 8}A_i A_i -{\mu \over 2}\epsilon_{ij}\Pi _i A_j 
]$$
\begin{equation}
 -p_2\eta ^2-q^2(\partial _i\Pi _i+{\mu\over 2}\epsilon_{ij}
\partial _iA_j )
-\bar C_1
\{\eta ^1,\bar\chi _1\}C^1 -\bar C_2
\{\eta ^2,\bar\chi _2\}C^2.
\label{41}
\end{equation}
The condition $m^2=0$ removes some of the terms in (\ref{01})
directly. The term ${{{\Pi _\psi }^2}\over{2m^2}}$ dominates over
rest of the $\Pi _\psi $ terms and is decoupled. Also the constraints
are appropriatly modified. Notice that now we have used both,
$$\int d^3x(p_1\dot q^1+\bar C_1\dot P^1)=\{Q_{BRST},\int d^3x~\bar C_1
\dot q^1\},$$
$$\int d^3x(p_2\dot q^2+\bar C_2\dot P^2)=\{Q_{BRST},\int d^3x~\bar C_2
\dot q^2\},$$
to remove the left hand side terms from the action. The gauge
$\eta ^1\equiv A_0$ is retained but $\eta ^2$ is kept arbitrary.

Clearly our first order Lagrangian now has the desired structure identical
to (\ref{97})
where the multipliers $\lambda _i$'s are to be identified with $q_2$ and
$p^2$ and $\Phi _l$s are 
$\eta ^2\approx 0$ and 
$\partial _i\Pi _i+{\mu\over 2}\epsilon_{ij}
\partial _iA_j \approx 0$. So far there is no change in the 
symplectic structure since the kinetic part of the Lagrangian has 
 retained its canonical form.
 According to the Faddeev-Jackiw procedure \cite{fj} one can
now use the "true" constraints $\Phi ^i$s in the theory to reduce the 
number of degrees of freedom. However, the corresponding changes in the
kinetic energy part of the action can induce a modification in the 
symplectic structure, to be computed as before. Notice that we have 
already done this job partially in using the constraints $A_0\approx 0 $
 and $\Pi ^0\approx 0$ strongly, which however 
does not lead to any change in the remaining brackets.

Indeed, according to \cite{fj}, it is not imperative to classify the 
constraints $\Phi _i$ according to FCC or SCC, or to compute the 
appropriate Dirac brackets. But from the BRST formalism we know that here
$\Phi _i \equiv (\chi _2,\eta ^2)$ constitute an SCC pair
since $\eta ^2$ is present in the gauge fixing fermion. Hence, although
$\eta ^2$ is arbitrary, we have to choose it such that 
$\{\chi _2,\eta ^2\}$ is non-vanishing.

Let us now fix the gauge $\eta ^2$ as,
\begin{equation}
\eta ^2\equiv \partial _i\Pi _i -{\mu \over 2}\epsilon _{ij}\partial _iA_j+
\epsilon _{ij}\partial _ih_j,
\label{42}
\end{equation}
where $h_j$ has a representation such that 
the combination $\epsilon _{ij}\partial _ih_j$ gives a non-trivial
contribution in $\{\chi _2,\eta ^2\}$. This choice reduces the
action of our model to, 
$$
{\cal L}=\epsilon ^{ij}({{\mu }\over 2}A^j-h^j)\dot A^i
-{1\over 2}h^i h^i-{1\over 2}(\epsilon ^{ij}\partial ^iA^j)^2 -q^2
(\mu \epsilon ^{ij}\partial ^iA^j -\epsilon ^{ij}\partial ^ih^j )$$
\begin{equation}
-\bar C_1
\{\eta ^1,\bar\chi _1\}C^1 -\bar C_2
\{\eta ^2,\bar\chi _2\}C^2.
\label{100}
\end{equation}
It should be mentioned that in the action 
 (\ref{41}) we are actually using $\Pi^i=
\epsilon ^{ij}({\mu \over 2}A^j-f^j)$ as a solution of (\ref{42}) and 
have not considered terms of the form $\epsilon^{ij}\partial ^j\alpha $.
This is slightly more restrictive than the gauge choice (\ref{42}).
The above action is immedietly recognisable as the master Lagrangian in
\cite{dj} once the fields are identified as $h_i\equiv f_i$ and $-q^2
\equiv A_0$. The ghost contribution remains decoupled so long as the
 gauge choices are linear.
The non-canonical symplectic structure also follows accordingly. The FCC
removes one degree of freedom leading to the single helicity mode of the
MCS theory. This completes the reduction of our system to that of the 
master Lagrangian constructed by Deser and Jackiw \cite{dj}.

\vskip 1cm
\begin{center}
{\bf VI: Conclusion}
\end{center}
\vskip .5cm

The power of the Batalin-Tyutin (BT) quantization scheme has been amply
demonstrated in the present work, where vector field theories in
2+1-dimensions have been considered. The models discussed primarily
are Maxwell-Proca (MP), Maxwell-Chern-Simons (MCS) and 
Maxwell-Chern-Simons-Proca (MCSP)
models. Of this group, the second one is a gauge theory and parity is
broken in the last two models due to the Chern-Simons term. The parity
violation is reflected by the fact that MCS theory generates a {\it 
{single}}
helicity mode and MCSP consists of {\it {two}} opposite helicity
 modes with {\it {unequal}} masses. The extra mode appears due to the
absence of gauge symmetry. This can be compared with MP model, where
the two helicities carry the {\it {same}} mass. The subtleties in
computing the spin content was discussed in \cite{dj} for MCS theory. 
Similar analysis were carried out in \cite{sub}, (an earlier
 version of the present work), and \cite{bkm} for
MP and MCSP models respectively. The present work shows how {\it {all}}
the above models can be discussed in the BT extended framework.

In the enlarged phase space, the BT extension of MCSP model has two
FCCs in involution. This system has been used in the (Hamiltonian) 
BRST quantization. The importance of the gauge choice has been 
demonstrated in relating various models. We have explicitly shown
that our BT extended model is equivalent to the conventional
Stuckelberg extension in a particular gauge, whereas in another gauge
it relates to the master Lagrangian discovered by Deser and Jackiw
 \cite{dj} to show the equivalence between self-dual model and
MCS theory. In our study of the latter connection, we have used
the symplectic quantization approach, initiated be Faddeev and
Jackiw \cite{fj}. All of the above discussion underlines the
significance of the master Lagrangian technique, which connects
apparantly different models and brings greater insight from their
equivalence.

\vskip 1cm

{\bf Acknowledgment:} It is a pleasure to thank Dr. Rabin Banerjee for
 fruitful discussions.

\end{document}